\documentclass[preprint]{aastex}
\usepackage{amsmath,amsfonts,amsthm,amssymb}
\usepackage{subfigure}
\usepackage{url}
\usepackage{rotating}

\newcommand{\R}{\mathbf{r}}

\newcommand{\bc}{\mathbf{b}}

\newcommand{\da}{\triangleq}

\def\mf{\mathbf}

\begin{document}

\title{Simulation and analysis of sub-$\mu$as precision astrometric data for planet-finding}

\author{Dmitry Savransky and N. Jeremy Kasdin}
\affil{Department of Mechanical and Aerospace Engineering\\ Princeton University, Princeton, NJ 08544}

\email{dsavrans@princeton.edu}

\begin{abstract}
We present a vector formulation of an interferometric observation of a star, including the effects of the barycentric motion of the observatory, the proper motions of the star, and the reflex motions of the star due to orbiting planets.  We use this model to empirically determine the magnitude and form of the signal due to a single Earth-mass planet orbiting about a sun-mass star.  Using bounding values for the known components of the model, we perform a series of expansions, comparing the residuals to this signal.  We demonstrate why commonly used first order linearizations of similar measurement models are insufficient for signals of the magnitude of the one due to an Earth-mass planet, and present a consistent expansion which is linear in the unknown quantities, with residuals multiple orders of magnitude below the Earth-mass planet signal.  We also discuss numerical issues that can arise when simulating or analyzing these measurements.
\end{abstract}


\section{Introduction}

Much study has been dedicated in recent years to the possibility of using an ultra-precise, space-based interferometer for the purpose of discovering extra-solar planets by their impact on the astrometric positions of their parent stars.  This has become one of the major science areas of the proposed Space Interferometry Mission (SIM) \citep{sozzetti2002,sozzetti2003,catanzarite2006}, and has also been considered as an application for the European Space Agency's Gaia mission \citep{casertano1996astrometry}.  Of particular interest to the exoplanet community is the possibility that interferometers capable of sub-$\mu$as precision can be used to detect the presence of Earth-sized planets in Earth-like orbits---a goal which is many years away from being realized by any of the other currently employed or studied planet-finding methods.  A number of studies have been completed in order to assess the exact planet-finding capabilities of astrometric instruments \citep{traub2009,casertano2008double,brown2009}.  One byproduct of these studies has been the realization that the classical description of astrometric observations (as described, for instance, in \citet{green1985spherical}) makes approximations that are suitable only when dealing with levels of precision of 1 mas or higher.  Several more precise descriptions have been published, including a very thorough one in \citet{konacki2002frequency}, but most of these take the classical approach of separately treating the effects of proper motion, parallax, and the stellar reflex due to companions, with separate expansions of each effect.  Furthermore, when demonstrating analysis techniques, these studies often still only use a first order expansion to simplify the required computations.  While it is possible to achieve the required numerical precision with these approaches, there is an added burden from having to separately consider the expansion of the direction vector and other effects.  We believe that a simpler approach is to linearize a single measurement equation to produce one unified expression.

Here, we derive the exact\footnote{By exact we do not mean that all possible contributions to the measurement are included; for instance, we have not yet considered relativistic effects.  Rather, we mean that we are formulating the exact form of the nonlinear measurement for the given set of effects included: parallax, proper motion, and stellar reflex.} expression for an astrometric measurement, and then present several expansions to multiple levels of precision.  This exercise is important for two reasons.  First, if one wishes to evaluate an algorithm, it is crucial to ensure that any simulated test data does not contain biases or components not present in the true data stream.  Even if such structures are below the level of other simulated noise sources, they may have an effect on any processing algorithm which makes the assumption of white, gaussian (or pseudo-gaussian) noise.  The added signals will not be random, and, as shown below, may closely resemble the signal sought in planet-finding applications.  For these reasons, we believe that the correct way to simulate astrometric data is to use an exact representation of the physical system being modeled.  This removes the possibility of inadvertently introducing non-random noise sources, or otherwise creating an unfair test for the analysis algorithm.  Second, when analyzing astrometric data, while linearization of the signal is a very useful tool, we must always ensure that such manipulation does not produce a template that is measurably different from the data.  If data is generated using the same linearization as is assumed by the analysis method, and the linearization introduces measurable structure not present in the true signal being simulated, then use of the same linearization in both simulation and analysis does not constitute a fair test of the algorithm.  Therefore, the main focus of this paper will not be a specific analysis technique.   Rather, we seek to  develop an exact formulation of the astrometric measurement so that completely unbiased data can be produced, on which various analysis techniques can be tested, and to examine the simplifications that can safely be made to this exact form.

\section{The astrometric measurement}\label{sec:measurement}

\begin{figure}[ht]
\centering
\includegraphics[width=0.85\textwidth]{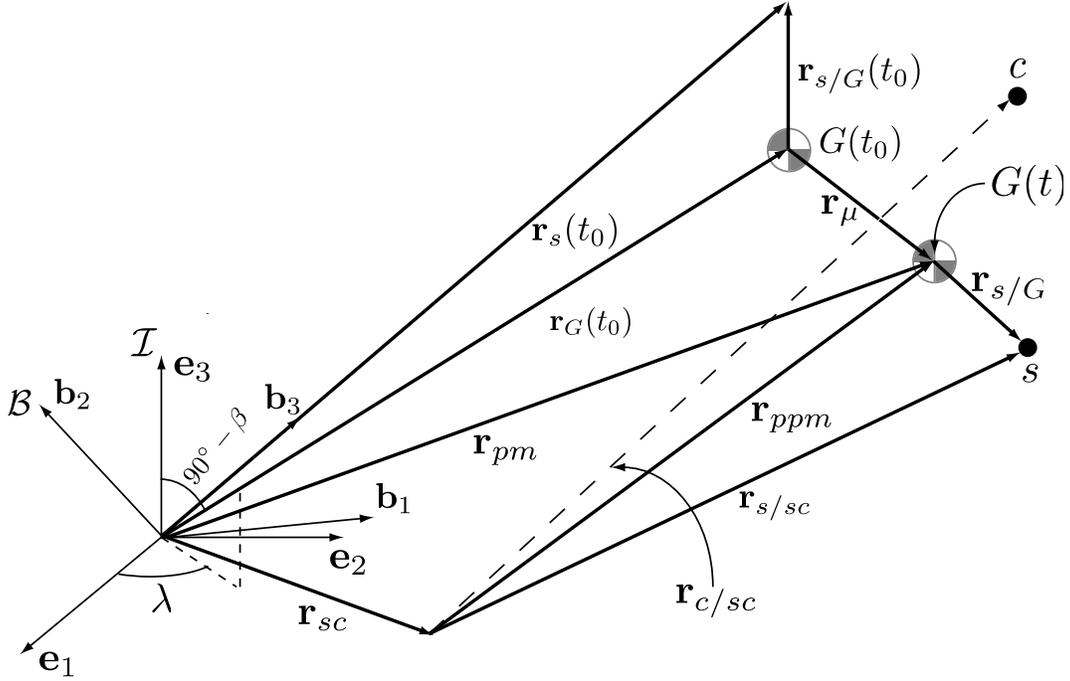}
 \caption[]{ A schematic of an astrometric observation.  The solar system barycenter is placed at the origin of frame $\mathcal{I}$, while the observed star system's barycenter at epoch $t_0$ is at $G(t_0)$ and at $G(t)$ at observation time $t$.  Point $s$ represents the star position at time $t$ and point $c$ represents the (possibly time-varying) position of the centroid of a group of reference stars.}
\label{fig:model} 
\end{figure}

Figure \ref{fig:model} and Table \ref{tbl:vectors} define the vectors and reference frames used to describe an astrometric observation of an exosystem.\footnote{In general, we will use the notation $\hat{\mathbf{x}}$ to denote the unit vector of $\mathbf{x}$ ($\hat{\mathbf{x}} =\mathbf{x}/\Vert \mathbf{x} \Vert$).  All bold symbols refer to vectors and terms without explicit time dependence are assumed to refer to time $t$.  We use $G$ to refer to the barycenter of the exosolar system (target star and planets).  The proper motion is defined as a motion of $G$.}
Reference frame $\mathcal{I}$ is an inertially fixed, barycentric/ecliptic frame located at the solar system barycenter with the unit vector $\mathbf{e}_3$ directed perpendicular to the ecliptic plane (the $\mathbf{e}_1$ and $\mathbf{e}_2$ unit vectors are arbitrary).  We define a second inertially fixed frame, called the {\em tangent frame}, $\mathcal{B}$, with $\mathbf{b}_3$ axis aligned with $\hat{\R}_s(t_0)$, the unit vector to the star at the initial time $t_0$.  The $\mathbf{b}_1$ axis is perpendicular to the plane containing $\hat{\R}_s(t_0)$ and $\mathbf{e}_3$.  The final unit vector direction, $\mathbf{b}_2$, is mutually perpendicular to $\bc_1$ and $\hat{\R}_s(t_0)$.
Given these definitions, we can find the ecliptic coordinates in $\mathcal{I}$ of the three unit vectors defining $\mathcal{B}$,
\begin{eqnarray}
\hat{\R}_s(t_0) = \bc_3 &=&  \left[ \begin{array}{ccc} \cos \lambda \cos \beta &\sin \lambda \cos \beta & \sin \beta \end{array}\right]^T_\mathcal{I} \label{eq:rhat_0def}\\
\bc_1 &=& \left[ \begin{array}{ccc} 0 & 0 & 1\end{array}\right]^T_\mathcal{I} \times \hat{\R}_s(t_0)/ \cos\beta =  \left[ \begin{array}{ccc} -\sin \lambda &\cos \lambda & 0 \end{array}\right]^T_\mathcal{I} \\
\bc_2 &=& \hat{\R}_s(t_0) \times \bc_1  = \left[ \begin{array}{ccc} -\cos \lambda \sin \beta &-\sin \lambda \sin \beta& \cos \beta \end{array}\right]^T_\mathcal{I}
\end{eqnarray}
where $\hat\R_s(t_0)$ is the unit vector to the star as given by its ecliptic coordinates at epoch $t_0$: $(\lambda,\beta)$, measured in the $\mathcal I$ frame.  This definition of frames assumes constant relative velocities between exosystem and local barycenters (i.e., constant proper motions).  For the time scales on which astrometric observations are taken, this is reasonable, but this derivation may have to be expanded if dealing with an accelerating exosystem.

We define the parallax by the small quantity,
\begin{equation}
\varpi \triangleq \frac{a}{\Vert \R_s(t_0) \Vert}
\end{equation}
where $a$ is a distance constant such that $\varpi$ is in units of radians.  Thus, $\R_s(t_0) = (a/\varpi) \hat{\R}_s(t_0)$ for $a = 1$ AU, with $\R_s(t_0)$ expressed in AU.  An estimate of $\varpi$ provides the distance to the target star at epoch $t_0$.  For a typical target star at 10 pc (2.0626$+\times10^6$ AU)\footnote{In the simulations in this paper, the conversion between pc and AU is always taken exactly, i.e., 1 pc = $\left(\tan\left(\frac{\pi}{180\times3600}\right)\right)^{-1}$ AU.}, $\varpi \sim 5 \times 10^{-7}$. The classical astrometric equations retain terms only to first order in $\varpi$ although, as we show in \S\ref{sec:expansion}, second order terms can be significant when working with $\mu$as precision measurements.

The motion of the barycenter of the target star system, $\R_\mu$, is considered as motion in the tangent reference frame,
\begin{equation}
\R_\mu(t) = \sigma_x (t - t_0) \bc_1 + \sigma_y (t - t_0) \bc_2 + \sigma_z (t - t_0) \hat\R_s(t_0)
\end{equation}
where $\sigma_i$ are the components of barycenter velocity at epoch $t_0$ in units of distance/time.  We approximate this velocity to be constant, as is usually done when considering short time spans such as  a space-based observatory's lifetime.  This expression is generally split into two components: the transverse and radial velocities.  Following \citet{green1985spherical}, we can write:
\begin{equation}\label{eq:veldef}
\begin{array}{c c l}
 \frac{{}^\mathcal{I} d}{dt} \R_G(t) &\equiv&  \frac{{}^\mathcal{I} d}{dt} \left(\R_G(t_0) + \R_\mu(t)\right) = \frac{{}^\mathcal{I} d}{dt}\R_\mu(t)\\
  &=& V_R  \hat\R_s(t_0) + \mathbf{V}_T \quad\textrm{where} \quad  \mathbf{V}_T  = \hat \R_s(t_0) \times \left(\frac{{}^\mathcal{I} d}{dt} \R_G(t) \times \hat \R_s(t_0) \right)
 \end{array}
\end{equation}
where the superscript on the derivative denotes differentiation in the inertial frame, and $\R_G$ is the position of the target system barycenter (see Figure \ref{fig:model}).  In our notation, $V_R$ is just $\sigma_z$, whereas $\mathbf{V}_T$ is the vector [$\sigma_x$, $\sigma_y$, 0]$^T$.  The transverse velocity (multiplied by time) thus gives the proper motion of the barycenter (motion in the plane of the sky), which has the largest effect on the position of the target system at the time of observation.  However, the radial velocity also has a measurable effect on the astrometric observation.  Because motion of the target system in the radial direction causes the distance to the system to change, we observe a difference in the direction to the target star, known as `perspective acceleration'.  As will be discussed in \S\ref{sec:expansion}, the third component of $\R_\mu$ (and thus $\sigma_z$) interacts in non-negligible ways with other values in the measurement, and is thus observable.  However, when dealing with a priori estimates of these velocity components, as in \S\ref{sec:expand_apriori}, it is important to note that the transverse and radial velocities are measured in different ways (i.e., astrometry vs. doppler spectroscopy).

We will find it convenient to have an expression for the normalized barycenter motion,
\begin{equation}
\bar{\R}_\mu \da \frac{\R_\mu}{\|\R_s(t_0)\|} = \bar \sigma_x (t - t_0) \bc_1 + \bar\sigma_y (t - t_0) \bc_2 + \bar\sigma_z (t - t_0) \hat{\R}_s(t_0)
\end{equation}
where $\bar\sigma_x$, $\bar\sigma_y$, and $\bar\sigma_z$ are the nondimensional barycenter velocities in radians/time unit. For typical stars considered as candidates for astrometric planet finding, angular velocities from 100 to 1000 mas are common, making $\bar\R_\mu$, over a 5-10 year period, of order $5 \times 10^{-7}$ to $5 \times 10^{-6}$.

Finally, we must consider what exactly is measured during an astrometric observation.  An interferometer measures the projection of a target direction onto the instrument's baseline vector, recorded as the optical path-length delay (OPD) between the two interferometer detectors,
\begin{equation}
\textrm{OPD} = \mf B \cdot \hat{\mf r}_{s/sc}  + k + n
\end{equation}
where $\mf B$ is the orientation vector of an interferometer of baseline length $B = \Vert \mf B \Vert$, $k$ is a constant term representing the offset of the optical path differences, and $n$ is the measurement noise.
Planet-finding is generally proposed in a narrow angle mode, where the measurement is the relative OPD between two sources in a field of view, performed in quick succession such that $k$ remains nearly constant, and is assumed to cancel in subtraction.  Typically each target star has multiple reference sources, so the combined differential OPD ($\Delta$OPD) is with respect to a centroid position.  The measurement is also repeated for two (preferably orthogonal) interferometer baseline orientations to track the 2D position of the target in the plane of the sky.  Assuming that the interferometer baselines are taken as directions $\mf b_1$ and $\mf b_2$ of our body frame, then the astrometric measurement becomes:
\begin{equation}\label{eq:measurement}
\mf d = B\left[ \begin{array}{l} \mf b_1 \cdot \left(\hat{\mf r}_{s/sc} - \hat{\mf r}_{c/sc}\right) \\\mf b_2 \cdot \left(\hat{\mf r}_{s/sc} - \hat{\mf r}_{c/sc}\right) \end{array}\right] + \mf n
\end{equation}
where $B$ is the size of the interferometer baseline and $\mf n$ is an additive noise vector due to measurement error.  

In the literature, the measurement in equation (\ref{eq:measurement}) is often described as the angular separation between the two sources, projected onto the baseline \citep{konacki2002frequency,sozzetti2002,sozzetti2003}; it is assumed that $\mf d$ scaled by $B$ is a radian measure that can be converted to other angular units such as arcseconds.  It is this assumption that leads to the terminology of $\mu$as-precise astrometry, as the required sensitivity of the instrument to changes in the OPDs, when treated as an angle, evaluates to under 1 $\mu$as.  

In fact, \citet{colavita1994measurement} points out that, using the definition for $\hat{\mf r}_{s/sc}$ in equation (\ref{eq:rhat_0def}), for a centroid separation of ($\Delta \lambda$, $\Delta \beta$), the difference in unit vectors, to first order, can be written as,
\begin{equation}\label{eq:sdiff_fo}
\hat{\mf r}_{s/sc} - \hat{\mf r}_{c/sc} \approx \left[\begin{matrix} 
\sin\beta\cos\lambda\Delta\beta + \cos\beta\sin\lambda\Delta\lambda\\
\sin\beta\sin\lambda\Delta\beta - \cos\beta\cos\lambda\Delta\lambda\\
-\cos\beta\Delta\beta \end{matrix}\right]_\mathcal{I} \, .
\end{equation}
In this way, knowledge of the baseline vector and the differential OPD allows you to calculate a vector which maps in a relatively simple fashion (to first order) to two spherical angles representing the separation between the target and centroid.

\begin{figure}[ht]
\centering
\includegraphics[width=0.4\textwidth]{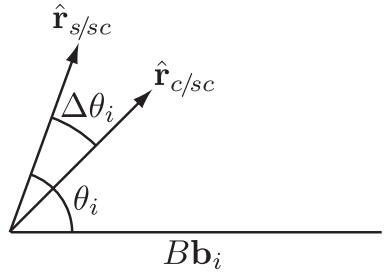}
 \caption[]{ A schematic of a narrow-angle astrometric measurement.  $B$ is the size of the interferometer baseline.}
\label{fig:narrow_mode_model} 
\end{figure}

Unfortunately, such a direct first order mapping between equation (\ref{eq:measurement}) and the angle between the target and centroid is not very accurate.  An OPD has units of distance or time (with the converting factor equal to the speed of light), and so a differential OPD will also have values equal to fractions of the interferometer baseline.  Let us assume that the (flat) wavefront from the target star is incident on the interferometer baseline with angle $\theta_i$, and that the centroid is located at an angle $\Delta\theta_i$ from the target star, in the projection of the sky due to interferometer orientation $\mf b_i$ (Figure \ref{fig:narrow_mode_model} --- $\Delta\theta_i$ is analogous to $\Delta\lambda$ and $\Delta\beta$ in equation (\ref{eq:sdiff_fo})).
The components of the measurement (without the noise term) are then,
\begin{equation}
d_i =B \left(\cos\theta_i - \cos(\theta_i - \Delta\theta_i)\right) =B\left(\cos\theta_i(1 - \cos\Delta\theta_i) - \sin\theta_i \sin\Delta\theta_i\right) \,.
\end{equation}

We are primarily interested in changes in the angle $\Delta\theta_i$, since this determines the movement of the target star with respect to the centroid in time (of course, this is further complicated by possible motion of the centroid itself).  Assuming this to be a small angle, we can substitute the Taylor series expansions of the sine and cosine terms in $\Delta\theta_i$ to first order to find,
\begin{equation}
d_i \approx -B \sin\theta_i \Delta\theta_i \, .
\end{equation}
Thus, by scaling the differential OPD by the interferometer baseline length and the direction on the sky of the target ($B\sin\theta_i$), we do get a first order approximation of the angular difference between the target and centroid (assuming perfect a priori knowledge of the target's (or equivalently centroid's) location with respect to the interferometer orientation).  Unfortunately, if we assume the target star to be on the order of 1$^\circ$ from the centroid \citep{sozzetti2002,shao1992potential}, the first of the dropped terms ($\Delta\theta_i^2/2$) has a magnitude of 1.5$\times10^{-4}$ rad, or 31.4 arcseconds.  When discussing ultra-precise applications, it is therefore inaccurate to treat the values produced by equation (\ref{eq:measurement}) as directly mapping to angular measures.  For these reasons, we treat all derived values as dimensionless for the remainder of this discussion, normalizing all distances and converting all angles to radians to remain consistent.  It is also important to point out that in this discussion, we consistently assume zero pointing error.  For a real instrument, accurate knowledge of the baseline vectors will be built up over many individual observations, each with an associated measurement error, but neither the interferometer orientation, nor the exact length of the baseline, can ever be known with perfect precision.  In order to include a pointing and baseline length error, we would need to extend the measurement equation, and update all subsequent calculations.

With these definitions and considerations, we can write the exact astrometric measurement  in terms of known quantities and the quantities whose values we wish to determine.  This is done by writing the vector from the spacecraft to the target star, $\mf r_{s/sc}$, in terms of the initial reference vector, $\R_s(t_0)$,
\begin{eqnarray}
\R_{s/sc} & = & \R_s(t_0) - \R_{s/G}(t_0) + \R_\mu + \R_{s/G} - \R_{sc} \nonumber \\
& = & \R_s(t_0) + \R_\mu + \Delta \R_{s/G} - \R_{sc} \,.
\end{eqnarray}
where $\R_{sc}$ is the spacecraft position vector relative to the solar system barycenter and  $\Delta \R_{s/G}$ is the difference in the star's position relative to $G$ between $t_0$ and epoch. We can then find the unit vector in the direction of $\R_{s/sc}$,
\begin{eqnarray}
\hat\R_{s/sc} &=& \frac{\R_{s/sc}}{\| \R_{s/sc} \|} \nonumber \\
&=& (\R_s(t_0) + \R_\mu +\Delta \R_{s/G}- \R_{sc})\times \nonumber \\
& & \left [\begin{array}{c}
 \R_s(t_0)\cdot\R_s(t_0) + \R_\mu \cdot \R_\mu + \Delta \R_{s/G} \cdot \Delta \R_{s/G} +\R_{sc} \cdot \R_{sc} + 2\R_s(t_0) \cdot \R_\mu \\
 + 2 \R_s(t_0) \cdot \Delta \R_{s/G} - 2\R_s(t_0) \cdot \R_{sc} + 2 \R_\mu \cdot \Delta \R_{s/G} - 2 \R_\mu \cdot \R_{sc} - 2 \Delta \R_{s/G} \cdot \R_{sc}
 \end{array}\right ]^{-\frac{1}{2}}.
 \label{eq:rhat_ssc}
\end{eqnarray}

The spacecraft to centroid pointing unit vector, $\hat{\mf r}_{c/sc}$, can be similarly expressed by using equation (\ref{eq:rhat_ssc}) for each reference source with the appropriate values for $\R_s(t_0)$ and $\R_\mu$.  When using this model for the reference sources, it is important to note that for a source $n$, $\hat\R_{n}(t_0)$ is not aligned with $\hat\R_s(t_0)$ and is thus not orthogonal to $\mf b_1$ and $\mf b_2$.  When evaluating equation (\ref{eq:measurement}), this simply means that all vectors have to be expressed as components in the same reference frame, which requires a change of coordinates for the reference sources.  This does, however, lead to some complications when dealing with linearizations of the model, which will be discussed further in \S\ref{sec:expansion}. If we assume extragalactic references with no companions or planets (which was done for the simulations described in this paper) the variation in the centroid position will be many orders of magnitudes below our desired level of accuracy.  Unfortunately, such references are not always available, and so these effects may become significant if one or more of the references is within our galaxy and has companions of its own.

When simulating an astrometric data set, it seems natural to use the formulation in equation (\ref{eq:rhat_ssc}).  The effects of parallax and proper motion are represented exactly, with no ambiguities in how perspective acceleration should be introduced.  Furthermore, the coupling of these effects with the astrometric wobble due to planetary systems also appears and does not need to be separately considered.  The computational effort required by this equation is not significantly greater than that of the various measurement expansions commonly in use, and is actually less than that of the second order expansions presented in the next section.  Other than a few numerical considerations discussed in \S\ref{sec:numerics}, the application of this formulation is trivial, and ensures high confidence in the simulated data.

\section{Measurement Expansion}\label{sec:expansion}
\begin{figure}[ht]
\centering
\includegraphics[width=0.75\columnwidth]{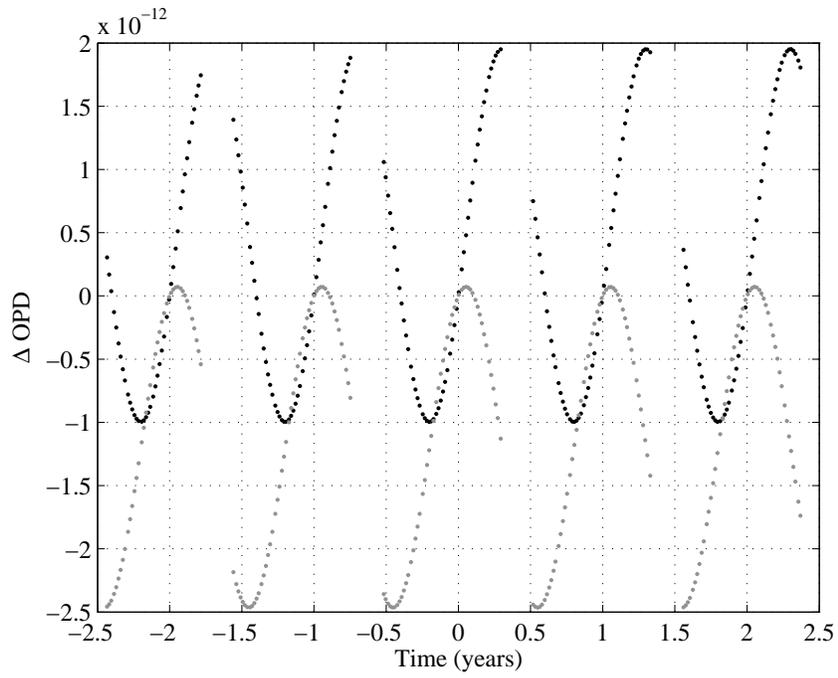}
 \caption{ Difference between differential OPDs (between a sun-twin star and fixed centroid) using exact expressions for $\hat\R_{s/sc}$ with and without the effects of an Earth-mass planet on an Earth-like orbit. The two curves represent the measurements along two orthogonal interferometer baseline orientations.\label{fig:planet_effects}}
\end{figure}

While equation (\ref{eq:rhat_ssc}) provides a method for exactly simulating precise astrometric measurements, the analysis of such signals can be made difficult by the nonlinear interactions of the various terms in the normalization factor.  The classical approach has been to expand this equation and retain only relevant terms.  As we will show below, for the purposes of planet-finding, these must necessarily include nonlinear terms, but an expansion is still useful as a simplifying step in the data analysis.  In order to decide which terms should be retained, we must first quantify the desired precision of the measurement.  There are several ways to do this, but a relatively straightforward one is to calculate the magnitude of the smallest signal we wish to be able to measure.  As our focus here is on the detection of exoplanets, we will take as our smallest signal the effect of a single Earth-mass planet on an Earth-like orbit on the astrometric signature of a sun-twin star. Figure \ref{fig:planet_effects} shows the differences between the results of two simulations, each using equation (\ref{eq:rhat_ssc}) and a fixed centroid position (with respect to the solar system barycenter).  The simulations include identical values for parallax, barycenter motion, and assume a sun-twin target star lying exactly 10 parsecs from the solar system barycenter (at epoch $t_0$). One simulation, however, includes the stellar jitter due to an Earth-mass planet, while the other does not (the line of nodes of the planet's orbit is rotated by 45$^\circ$ with respect to the line of sight).  The difference between the two measurements is the most basic representation of the magnitude of the signal in which we are interested, and, in this case, shows a signal on the order of 1$\times10^{12}$. This value is in the units of equation (\ref{eq:measurement}): it is the difference between fractions of the baseline distance. 

This tells us that to be sensitive to the jitter due to an Earth-twin with any measure of confidence, we must retain all terms in the measurement of order 1$\times10^{13}$ or greater (assuming that our analysis technique will be sensitive to structured signals up to one order of magnitude below the target signal).  To aid in this calculation, Table \ref{tbl:typ_vals} enumerates the ranges of typical values for the various terms in equation (\ref{eq:rhat_ssc}).

\subsection{Expansion Assuming No Prior Knowledge}\label{sec:expand_noapriori}
We begin the expansion by dividing the numerator and denominator of equation (\ref{eq:rhat_ssc}) by $\|\R_s(t_0)\|$ to find the normalized version,
\begin{eqnarray} \label{eq:rhat_norm}
\hat\R_{s/sc} &=& (\hat\R_s(t_0) + \bar\R_\mu +\varpi\Delta \tilde\R_{s/G}- \varpi\tilde\R_{sc})\times \\
& & \left [\begin{array}{c}
 1 + \bar\R_\mu \cdot \bar\R_\mu + \varpi^2\Delta \tilde\R_{s/G} \cdot \Delta \tilde\R_{s/G} +\varpi^2\tilde\R_{sc} \cdot \tilde\R_{sc}  + 2\hat\R_s(t_0) \cdot \bar\R_\mu \\
+ 2 \varpi \hat\R_s(t_0) \cdot \Delta \tilde\R_{s/G} - 2\varpi\hat\R_s(t_0) \cdot \tilde\R_{sc}  + 2\varpi \bar\R_\mu \cdot  \Delta \tilde\R_{s/G}  - 2\varpi \bar\R_\mu \cdot \tilde\R_{sc} - 2\varpi^2 \Delta \tilde\R_{s/G} \cdot \tilde\R_{sc}
 \end{array}\right ]^{-\frac{1}{2}}  \nonumber
\end{eqnarray}
where $\varpi = a/\|\R_s(t_0)\|$ and $\bar\R_\mu = \R_\mu/\|\R_s(t_0)\|$, $\tilde\R_{sc}$ is the spacecraft position normalized by $a$ and $\Delta \tilde\R_{s/G}$ is  $\Delta \R_{s/G}$, also normalized by $a$.  

We next apply a binomial expansion to the denominator of equation (\ref{eq:rhat_norm}) and retain terms explicitly in $\varpi$ to first order.  Since the signal we are looking for appears explicitly only as $\varpi\Delta \tilde\R_{s/G}$, any terms of proportional order must be retained.  Thus, terms proportional to $\Vert \bar\R_\mu \Vert^2$ must be left in, but terms proportional to $\Vert \bar\R_\mu \Vert^n$ for $n > 2$ and terms proportional to $\Vert \bar\R_\mu \Vert^n\varpi$ for $n > 1$ are dropped.  We also drop all terms proportional to $\hat\R_s(t_0)$, as dotting $\hat\R_{s/sc}$ with $\mf b_i$ as in the measurement equation (\ref{eq:measurement}) causes these terms to equal zero.  The resulting approximation is:
\begin{equation}\label{eq:rhat_expand1}
\hat\R_{s/sc}\cdot \mf b_i \approx
\left(\begin{array}{l}
 \bar\R_\mu +  \varpi \Delta \tilde\R_{s/G}+ \varpi (\hat\R_s(t_0) \cdot \tilde\R_{sc}) \bar\R_\mu - \varpi(\Delta\tilde\R_{s/G} \cdot \hat\R_s(t_0))\bar\R_\mu  - \varpi \tilde \R_{sc} \\
{}  - (\hat\R_s(t_0) \cdot \bar\R_\mu)\bar\R_\mu +\varpi  (\hat\R_s(t_0)\cdot\bar\R_\mu)\tilde \R_{sc}-\varpi(\hat\R_s(t_0)\cdot\bar\R_\mu)\Delta \tilde\R_{s/G}
\end{array}\right) \cdot \mf b_i \,.
\end{equation}
This is  a linear expression in parallax and can be used to extract the stellar reflex signal, though it contains complicated couplings among the spacecraft position, parallax, barycenter motion, and stellar reflex.  Note that it is not assumed that these other terms are known; rather, the barycenter motion and the parallax factor, $\varpi$, must also be estimated in the data analysis.  This leads to the problem of attempting to fit the radial motion of the target from purely astrometric measurements.  While the radial and proper motions of targets are approximated as completely separate in the classical treatment, the level of precision required for this application means that we must consider the effects of radial motion.  The first term of equation \ref{eq:rhat_expand1} is $\bar\R_\mu \cdot \mf b_i$.  If $\mf b_i$ is, as we have assumed, equivalent to $\mf b_1$ or $\mf b_2$, then the $\sigma_z$ dependence of this term is exactly zero.  Similarly, the next two terms in the direction of $\bar\R_\mu$ will have no $\sigma_z$ components when dotted with $\mf b_i$.  However, the two final terms in this expansion have magnitudes given by $\varpi  (\hat\R_s(t_0)\cdot\bar\R_\mu)$, which makes them proportional to $\sigma_z$.  As these two terms are in the directions of $ \tilde \R_{sc}$ and $\Delta \tilde\R_{s/G}$, which are arbitrarily oriented in the tangent frame, this $\sigma_z$ dependence is not zeroed even when the baseline is perfectly aligned with $\mf b_1$ or $\mf b_2$.  Thus, the radial star motion explicitly enters our expression, and makes a significant contribution at the desired level of precision.

\begin{figure}[ht]
\centering
\includegraphics[width=0.75\columnwidth]{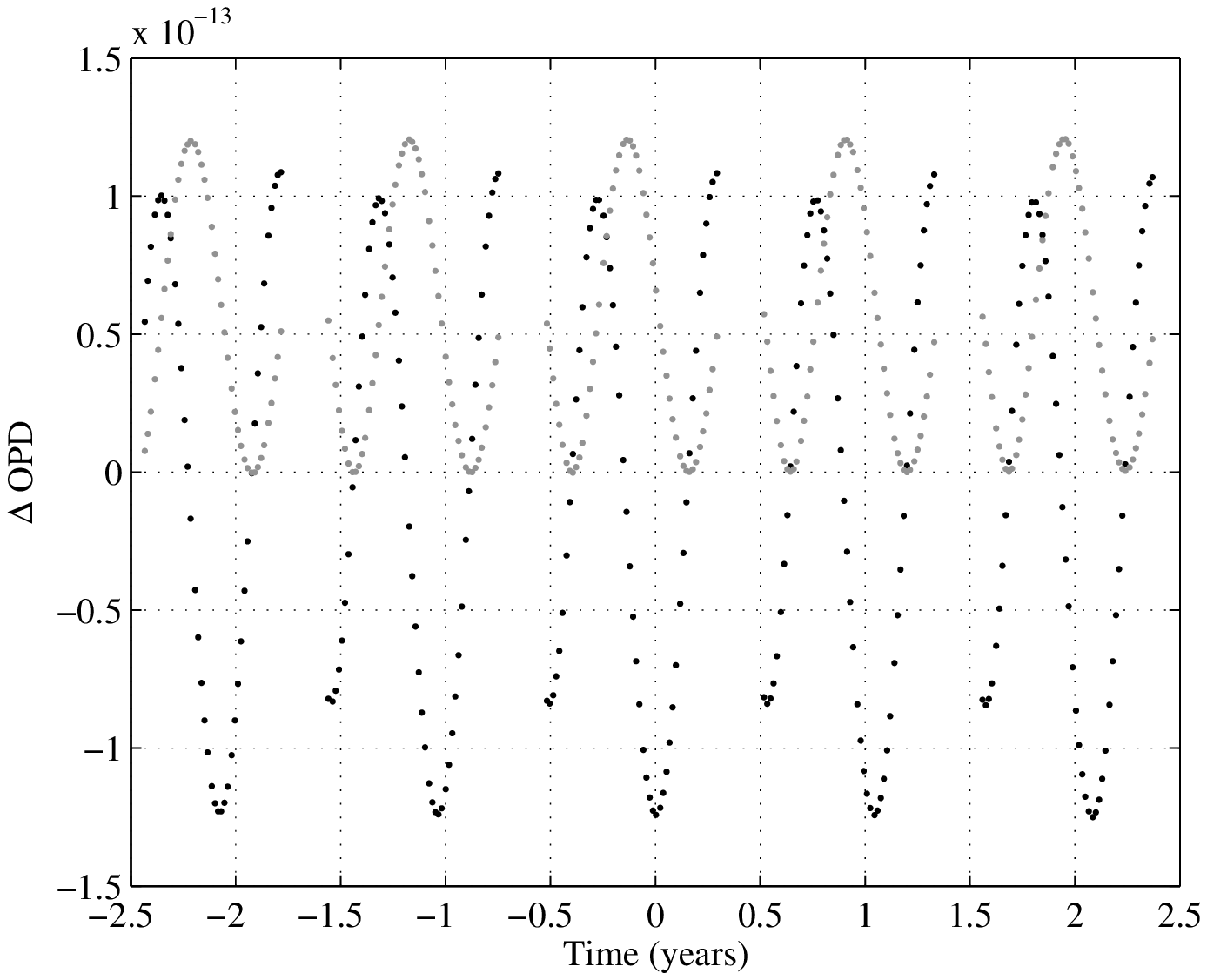}
 \caption[]{ Difference between differential OPDs (between a sun-twin star and fixed centroid) using the exact expression for $\hat\R_{s/sc}$ and the first order expansion from equation (\ref{eq:rhat_expand1}). The two curves represent the measurements along two orthogonal interferometer baseline orientations.\label{fig:first_order_expansion}}
 \end{figure}

\begin{figure}[ht]
\centering
\includegraphics[width=0.75\columnwidth]{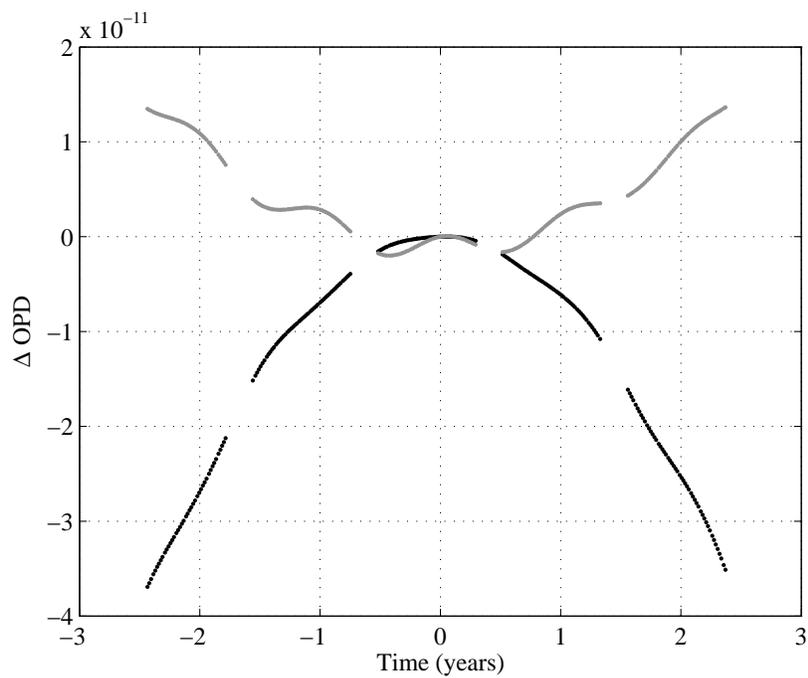}
 \caption[]{ Difference between differential OPDs (between a sun-twin star and fixed centroid) using the exact expression for $\hat\R_{s/sc}$ and the first order expansion from equation (\ref{eq:rhat_expand1}) with the radial terms neglected. The two curves represent the measurements along two orthogonal interferometer baseline orientations.\label{fig:first_order_expansion_norv}}
 \end{figure}

Figure \ref{fig:first_order_expansion} repeats the simulation from Figure \ref{fig:planet_effects}, including the planet in both cases, but now comparing the exact expression for $\hat\R_{s/sc}$ and the first order expansion in  equation \ref{eq:rhat_expand1}.  The resulting error is less than one order of magnitude below the signal due to the planet, which means that it is of the same order as the desired sensitivity for all targets closer than 10 pc or with any combination of the other parameters which would cause the planet signal to decrease in magnitude.  Additionally, the structure of the error is dominated by parallax terms, which produce periodic structure that could be mistaken for astrometric wobble due to another planet.  We can compare equation \ref{eq:rhat_expand1} with previously published first-order forms such as Equations 31 and 32 in \citet{konacki2002frequency}.  These have a very similar form to equation \ref{eq:rhat_expand1}, save that it is assumed that all radial components of motion are unobservable, and thus the equivalent vectors to $\Delta \tilde\R_{s/G}$ and $\bar\R_\mu$ are formed without the $\hat\R_s(t_0)$ components.  The expression there also does not subtract the initial displacement of the target star from its system's barycenter, thereby applying the proper motion to the star rather than the system barycenter.  If we redo our simulation using the first order expansion with the radial terms dropped, we end up with residuals one order of magnitude higher than those in figure \ref{fig:first_order_expansion}.  Thus, the exclusion of radial terms produces residuals that are one order of magnitude above the target signal, rather than one order of magnitude below, as seen in Figure \ref{fig:first_order_expansion_norv}.

Furthermore, a number of the terms dropped in this expansion were proportional to $\varpi^2$ multiplied by quantities of order 1.  The astrometric literature is quite clear that a first order expansion in $\varpi$ is only good to milliarcsec accuracy. \citep{green1985spherical} If our instrument has a final precision of sub-$\mu$arcsec, it is important to ask if measurable terms were dropped in the expansion.  Such terms can get as large as 6 times $\varpi^2$.  Again, for a star at 10 pc, $\varpi \sim 5 \times 10^{-7}$, making the second order terms of order $2.5 \times 10^{-13}$.   Thus, for most stars the approximation is a good one but it does raise concerns for the closest targets, or for analysis methods that are sensitive to non-random signals just below the level of the target signal.  

\begin{figure}[ht]
\centering
\includegraphics[width=0.75\columnwidth]{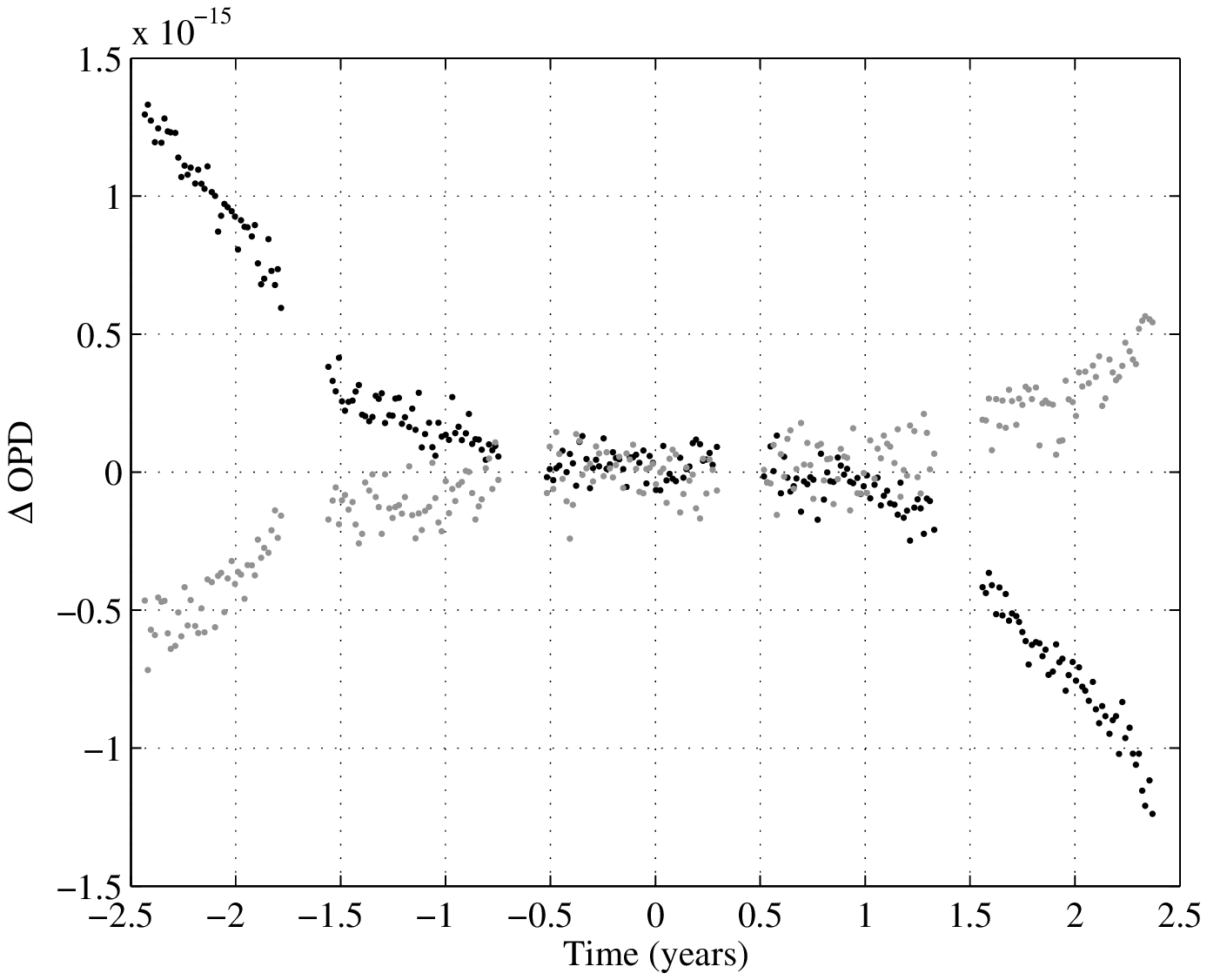}
 \caption[]{ Difference between differential OPDs (between a sun-twin star and fixed centroid) using the exact expression for $\hat\R_{s/sc}$ and the second order expansion from equation (\ref{eq:rhat_expand2}). The two curves represent the measurements along two orthogonal interferometer baseline orientations. \label{fig:second_order_expansion}}
\end{figure}

To address this, we can repeat the expansion to second order in $\varpi$.  However, for consistency, we must now retain some additional terms.  The zeroeth and first order terms (in $\varpi$) remain the same (since $\Vert \bar\R_\mu \Vert^3$ is of order 1$\times10^{-16}$ and  $\Vert \bar\R_\mu \Vert^2\varpi$ is of order 1$\times10^{-17}$).  Any terms in $\varpi^2$ with factors of $\bar\R_\mu$ can safely be dropped.  Terms in $\varpi^2$ with factors of $\Delta \tilde\R_{s/G}$ should also be quite small, but we will leave these to first order since it is theoretically possible that there exist systems with both Earth-like planets and very large, widely separated super-Jupiters which have not yet been detected by RV (due to their very long periods), and which would cause a large stellar reflex to make these factors significant.  These considerations leave only three new terms in the expansion:

\begin{equation}\label{eq:rhat_expand2}
\hat\R_{s/sc} \cdot \mf b_i \approx
\left(\begin{array}{l}
\bar\R_\mu  - (\hat\R_s(t_0) \cdot \bar\R_\mu)\bar\R_\mu +  \left(\Delta \tilde\R_{s/G}  - \tilde \R_{sc}  -(\hat\R_s(t_0)\cdot\bar\R_\mu)\Delta \tilde\R_{s/G}\right)\varpi\\
{} + \left( (\hat\R_s(t_0) \cdot \tilde\R_{sc}) \bar\R_\mu - (\Delta\tilde\R_{s/G} \cdot \hat\R_s(t_0))\bar\R_\mu +(\hat\R_s(t_0)\cdot\bar\R_\mu)\tilde \R_{sc}\right)\varpi\\
{} + \left(\tilde\R_{sc} (\hat\R_s(t_0) \cdot \Delta \tilde\R_{s/G}) + \Delta \tilde\R_{s/G} (\hat\R_s(t_0) \cdot \tilde\R_{sc}) - \tilde\R_{sc} (\hat\R_s(t_0) \cdot \tilde\R_{sc})\right)\varpi^2
\end{array}\right)  \cdot \mf b_i \, .
\end{equation}
Figure \ref{fig:second_order_expansion} shows the difference between the exact expression for $\hat\R_{s/sc}$ and the second order expansion in  equation (\ref{eq:rhat_expand2}).  The resulting error is now two to three orders of magnitude below the target signal.  The error structure is now dominated by proper motion terms, which are linear, not periodic.  These could still be confused with the effects of long period planets, but the signals are so low compared to the target signal that this possibility is unlikely.  It is important to note, however, the fitting problem has become significantly more challenging as the parameters enter nonlinearly in the measurement.

As has already been mentioned, our practice of dropping terms that are in the direction of $\hat\R_s(t_0)$ breaks down in two cases: when there is pointing error such that the interferometer baseline is not aligned with $\mf b_1$ or $\mf b_2$, or when expressing the direction to a different star (i.e., a non-ideal reference source) in the tangent frame defined by the target star.  In these cases, a number of additional terms must be included in our expansion.  Following the same procedure as above, we can rewrite equation (\ref{eq:rhat_expand2}) as:
\renewcommand{\arraystretch}{1.2}
\begin{equation}\label{eq:rhat_expand2_perr}
\hat\R_{s/sc} \approx
\left(\begin{array}{l}
\hat\R_s(t_0) + \bar\R_\mu - \hat\R_s(t_0) (\hat\R_s(t_0) \cdot \bar\R_\mu) - \bar\R_\mu (\hat\R_s(t_0) \cdot \bar\R_\mu)\\
{} -  \frac{\hat\R_s(t_0)}{2} (\bar\R_\mu \cdot \bar\R_\mu) + \frac{3\hat\R_s(t_0)}{2} (\hat\R_s(t_0) \cdot \bar\R_\mu)^2 \\
+ \left[\begin{array}{l}
\Delta \tilde\R_{s/G} - \tilde\R_{sc} - \Delta \tilde\R_{s/G} (\hat\R_s(t_0) \cdot \bar\R_\mu) + \tilde\R_{sc} (\hat\R_s(t_0) \cdot \bar\R_\mu)\\
 {}- \hat\R_s(t_0) (\hat\R_s(t_0) \cdot \Delta \tilde\R_{s/G}) - \bar\R_\mu (\hat\R_s(t_0) \cdot \Delta \tilde\R_{s/G})\\
 {} + 3\hat\R_s(t_0) (\hat\R_s(t_0) \cdot \bar\R_\mu) (\hat\R_s(t_0) \cdot \Delta \tilde\R_{s/G}) + \hat\R_s(t_0) (\hat\R_s(t_0) \cdot \tilde\R_{sc})\\
 {} + \bar\R_\mu (\hat\R_s(t_0) \cdot \tilde\R_{sc}) - 3\hat\R_s(t_0) (\hat\R_s(t_0) \cdot \bar\R_\mu) (\hat\R_s(t_0) \cdot \tilde\R_{sc})\\
 {} - \hat\R_s(t_0) (\bar\R_\mu \cdot \Delta \tilde\R_{s/G}) + \hat\R_s(t_0) (\bar\R_\mu \cdot \tilde\R_{sc})
\end{array}\right] \varpi  \\
+ \left[\begin{array}{l}
\tilde\R_{sc} (\hat\R_s(t_0) \cdot \Delta \tilde\R_{s/G}) + \Delta \tilde\R_{s/G} (\hat\R_s(t_0) \cdot \tilde\R_{sc}) - \tilde\R_{sc} (\hat\R_s(t_0) \cdot \tilde\R_{sc})\\
{} - 3\hat\R_s(t_0) (\hat\R_s(t_0) \cdot \Delta \tilde\R_{s/G}) (\hat\R_s(t_0) \cdot \tilde\R_{sc}) + \hat\R_s(t_0) (\Delta \tilde\R_{s/G} \cdot \tilde\R_{sc})\\
{} - \frac{\hat\R_s(t_0)}{2} (\tilde\R_{sc} \cdot \tilde\R_{sc})  + \frac{3\hat\R_s(t_0)}{2} (\hat\R_s(t_0) \cdot \tilde\R_{sc})^2 
\end{array}\right] \varpi^2
\end{array}\right)
\end{equation}
\renewcommand{\arraystretch}{1}
where the subscript $s$ may now refer to either the target star, or any of its reference sources.

We now have a completely general second order expansion of the spacecraft to source unit vector, which applies when the source is different from the target star used to define the tangent frame, or when the spacecraft baseline pointing does not coincide with the vectors of the tangent frame.  In the first case, we retain terms proportional to $\hat\R_s(t_0) \cdot \mf b_i$ because $\hat\R_s(t_0) \ne \mf b_3$ and so these terms do not automatically go to zero.  In the second case, these terms do not go to zero because $\mf b_i \ne \mf b_1$ or $\mf b_2$.  This expression is even more complex than equation  (\ref{eq:rhat_expand2}), and even less suitable for use in analysis.  However, this added complexity is actually mitigated by the same reasons that caused us to add in the extra terms.  First, if the pointing error is small (i.e., the baseline is reasonably close to one of the two tangent frame unit vectors), then only the new terms in $\hat\R_s(t_0)$ whose coefficients are of order 1 should be kept, as $\hat\R_s(t_0) \cdot \mf b_i$ will still represent a small value.  Second, even non-ideal reference sources will still usually be stars which are significantly further from the solar system than the target star, with much smaller parallax and barycenter motion values.  As such, the first order form will typically be sufficiently precise for these sources, allowing most or all of the terms in $\varpi^2$ to be dropped.

\subsection{Expansion With Prior Knowledge}\label{sec:expand_apriori}
While the expansions in \S\ref{sec:expand_noapriori} are a useful representation of an astrometric measurement, they do not quite correspond to the real-life analysis problem of astrometric planet-finding.  In fact, before any high precision astrometric measurements are made, we will already have some knowledge of the barycenter motion and parallax of our target stars, which will only need  to be updated by some small amount.  To represent this, we can rewrite equation (\ref{eq:rhat_norm})
with $\bar\R_{\mu}$ replaced by $\bar\R_{\mu0}+ \delta\bar\R_\mu$ and $\varpi$ replaced by $\varpi_0 + \delta\varpi$, where $\bar\R_{\mu0}$ and $\varpi_0$ are the currently known components of barycenter motion and parallax and $\delta\bar\R_\mu$ and $\delta\varpi$ are the (relatively small) correction terms. The result is:
\renewcommand{\arraystretch}{1.2}
\begin{eqnarray}
\hat\R_{s/sc} &=& (\hat\R_s(t_0) + \bar\R_{\mu0} + \delta\bar\R_\mu  +(\varpi_0 + \delta\varpi)\Delta \tilde\R_{s/G}- (\varpi_0 + \delta\varpi)\tilde\R_{sc})\times\label{eq:rhat_norm2}  \\
& & \left [\begin{array}{l}
 1 + \bar\R_{\mu0} \cdot \bar\R_{\mu0} + \delta\bar\R_\mu \cdot \delta\bar\R_\mu + 2 \bar\R_{\mu0} \cdot \delta\bar\R_\mu + (\varpi_0 + \delta\varpi)^2\Delta \tilde\R_{s/G} \cdot \Delta \tilde\R_{s/G} +(\varpi_0 + \delta\varpi)^2\tilde\R_{sc} \cdot \tilde\R_{sc}\\
 {}  + 2\hat\R_s(t_0) \cdot \bar\R_{\mu0} + 2\hat\R_s(t_0) \cdot \delta\bar\R_\mu + 2 (\varpi_0 + \delta\varpi) \hat\R_s(t_0) \cdot \Delta \tilde\R_{s/G} - 2(\varpi_0 + \delta\varpi)\hat\R_s(t_0) \cdot \tilde\R_{sc}\\
 {} +  2(\varpi_0 + \delta\varpi) (\bar\R_{\mu0}  \cdot  \Delta \tilde\R_{s/G} + \delta\bar\R_\mu \cdot  \Delta \tilde\R_{s/G} ) - 2(\varpi_0 + \delta\varpi) (\bar\R_{\mu0}\cdot \tilde\R_{sc} + \delta\bar\R_\mu\cdot \tilde\R_{sc}) \\
 {} - 2(\varpi_0 + \delta\varpi)^2 \Delta \tilde\R_{s/G} \cdot \tilde\R_{sc}
 \end{array}\right ]^{-\frac{1}{2}}
\nonumber
\end{eqnarray}
\renewcommand{\arraystretch}{1}

In order to carry out expansions as before, it is necessary to quantify the expected values of these new variables so that we can decide which terms are negligible and which are important.  Since we are assuming that our a priori estimates will be reasonably good, $\bar\R_{\mu0}$ and $\varpi_0$ should be of the same order as $\R_\mu$ and $\varpi$ above.  Currently, parallax and proper motion are known to at least 1 mas and 1 mas/year accuracy, respectively \citep{turon1995}.  We will take  $\delta\varpi$ to be of order 5$\times10^{-9}$; $\Vert\delta\bar\R_\mu\Vert$ will vary in magnitude in time, but could be of order 1$\times10^{-8}$ within five years of observations.  

We again perform a binomial expansion on the denominator of equation (\ref{eq:rhat_norm2}) and retain terms explicitly in $\varpi_0$ to second order and explicitly in $\delta\varpi$ to first order.   As before, terms proportional to $\Vert \bar\R_{\mu0} \Vert^n$ for $n > 2$ and terms proportional to $\Vert \bar\R_{\mu0} \Vert^n\varpi_0$ for $n > 1$ are dropped, as are terms proportional to  $\Vert \delta\bar\R_{\mu} \Vert^n$ for $n > 1$, terms proportional to $\Vert \delta\bar\R_{\mu} \Vert\delta\varpi$, and terms including $\varpi^2$ that are proportional to $\Vert\Delta \tilde\R_{s/G}\Vert^n$ for $n>1$.

Terms proportional to $\Vert \bar\R_{\mu0} \Vert \delta\varpi$, $\Vert \bar\R_{\mu0} \Vert\Vert \delta\bar\R_{\mu} \Vert$, and $\Vert \delta\bar\R_{\mu} \Vert\varpi_0$ are borderline---for the values assumed here, they have magnitudes of up to 1$\times10^{-14}$, which means they should be neglected, but if the proper motions are larger than expected (or there is higher uncertainty in the a priori measurements), then these terms can become significant, so we will leave them.  Similarly, terms proportional to $\varpi_0\delta\varpi$  have magnitudes of 1$\times10^{-15}$ for the values assumed here, and are fairly unlikely to become significant unless uncertainties in parallax are much greater than expected for the nearest stars.  We include only one of these terms because leaving it out leaves a clear sinusoidal pattern in the signal of order 5$\times10^{-15}$.  If the chosen analysis method is not expected to be sensitive to  such a signal, this term can safely be omitted.  The resulting approximation is:

\renewcommand{\arraystretch}{1.2}
\begin{equation} \label{eq:rhat_expand_apri}
\hat\R_{s/sc} \cdot \mf b_i \approx
\left(\begin{array}{l}
\bar\R_{\mu0} + \delta\bar\R_\mu - \bar\R_{\mu0} (\hat\R_s(t_0) \cdot \bar\R_{\mu0}) - \delta\bar\R_\mu (\hat\R_s(t_0) \cdot \bar\R_{\mu0}) - \bar\R_{\mu0} (\hat\R_s(t_0) \cdot \delta\bar\R_\mu)\\
{} + \left[\begin{array}{c} 
\Delta \tilde\R_{s/G} - \tilde\R_{sc} - \Delta \tilde\R_{s/G} (\hat\R_s(t_0) \cdot \bar\R_{\mu0}) \\
 {}+ \tilde\R_{sc} (\hat\R_s(t_0) \cdot \bar\R_{\mu0}) + \bar\R_{\mu0} (\hat\R_s(t_0) \cdot \tilde\R_{sc})
 \end{array}\right] \left(\delta\varpi +\varpi_0\right)\\
 {} - \bar\R_{\mu0} (\hat\R_s(t_0) \cdot \Delta \tilde\R_{s/G})\delta\varpi + \left[ \tilde\R_{sc} (\hat\R_s(t_0) \cdot \delta\bar\R_\mu) + \delta\bar\R_\mu (\hat\R_s(t_0) \cdot \tilde\R_{sc})\right] \varpi_0\\
 {}  -2\tilde\R_{sc} (\hat\R_s(t_0) \cdot \tilde\R_{sc})\varpi_0 \delta\varpi\\
 {} + \left[\tilde\R_{sc} (\hat\R_s(t_0) \cdot \Delta \tilde\R_{s/G}) + \Delta \tilde\R_{s/G} (\hat\R_s(t_0) \cdot \tilde\R_{sc}) - \tilde\R_{sc} (\hat\R_s(t_0) \cdot \tilde\R_{sc})\right] \varpi_0^2
\end{array}\right)\cdot \mf b_i
\end{equation}
\renewcommand{\arraystretch}{1}

\begin{figure}[ht]
\centering
\includegraphics[width=0.75\columnwidth]{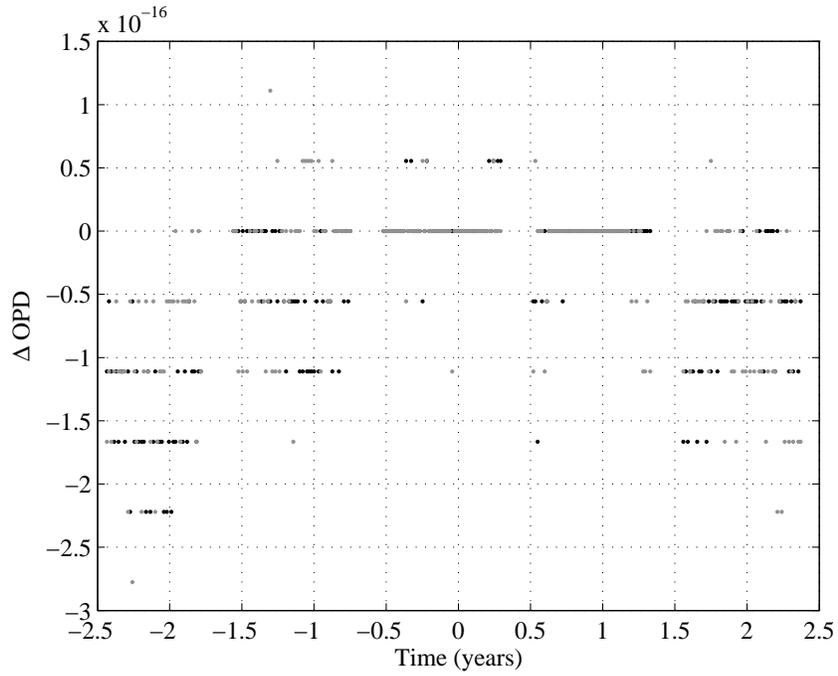}
 \caption[]{Difference between differential OPDs (between a sun-twin star and fixed centroid) using the second order expansion of $\hat\R_{s/sc}$ from equation (\ref{eq:rhat_expand2}) and the  expansion from equation (\ref{eq:rhat_expand_apri}) with assumed errors in parallax and barycenter motion of 1 mas and 1 mas/year, respectively.  The pattern in the residuals is due to the precision of the numerical data type used. The two curves represent the measurements along two orthogonal interferometer baseline orientations. \label{fig:apri_expansion}}
\end{figure}

The difference between this equation and the OPD using the exact  expression for  $\hat\R_{s/sc}$ is almost exactly the same as for the second order expansion from \S\ref{sec:expand_noapriori}.  In fact, the difference between differential OPDs calculated using equations (\ref{eq:rhat_expand2}) and (\ref{eq:rhat_expand_apri}) is actually at the level of the numerical precision of the data type used in these simulations (Figure \ref{fig:apri_expansion}).  However, unlike equation (\ref{eq:rhat_expand2}), this equation is linear in the quantities to be estimated, which can make analysis much simpler.

One assumption made in the above analysis is that the error is of the same magnitude for all three components of $\bar\R_{\mu}$.  However, this ignores the fact that currently existing estimates for the first two components (the transverse motion) are derived from different sources than estimates for the third component (the radial motion).  While proper motions have been measured by space-based astrometric instruments, radial velocities are generally provided by ground-based doppler spectroscopy, and are inherently less accurate.  To address this, we can modify equation (\ref{eq:rhat_expand_apri}) by separating $\delta\bar\R_\mu$ into two separate errors so that:
\begin{equation}\label{eq:rmu_redef}
\bar\R_{\mu} =  \bar\R_{\mu0}  +\delta\R_{R} + \delta\R_{T} 
\end{equation}
where $\delta\R_{R}$ and $\delta\R_{T}$ are due to errors in our prior knowledge of $V_R$ and $\mf V_T$, as defined in equation (\ref{eq:veldef}).

 We can now repeat the expansion performed to generate equation  (\ref{eq:rhat_expand_apri}), substituting the right-hand side of equation (\ref{eq:rmu_redef}) for $\bar\R_{\mu}$. The rules for keeping terms containing $ \delta\R_{T}$ remain the same as those for $\delta\bar\R_\mu$.  If we assume that $\delta\R_{R}$ is less than three orders of magnitude higher than $\delta\R_{T}$ (i.e., m/s radial velocity precision), then the expansion in equation (\ref{eq:rhat_expand_apri}) must also retain terms proportional to  $\Vert \delta\R_{R} \Vert^n$ for $n <= 2$.  The result is that only one term, equal to $-\delta\R_{R}( \hat\R_s(t_0) \cdot  \delta\R_{R}) = -\delta\R_{R}\Vert\delta\R_R\Vert$, must be added to equation (\ref{eq:rhat_expand_apri}) to account for the lower precision of the prior radial velocity measurement.

\section{Numerical Considerations}\label{sec:numerics}

The final simulation results presented in Figure \ref{fig:apri_expansion} highlight a very important point.  Due to the small magnitudes of many of the quantities utilized in this analysis, when it comes time to numerically manipulate them, the limitations of the numerical representation become a real concern.  Any computer generated value will have an associated finite precision, but specific representation schemas may actually introduce precision differences as functions of value magnitude.  For example, in MATLAB (which was used for the simulations in this paper), the default numerical data type is a double precision, floating point number, encoded as,
\begin{equation*}
2^e\times s
\end{equation*}
where $e$ is the exponent and $s$ is the significand.  These are stored as 11 bits and 53 bits, respectively, as per the IEEE 754 standard.  Because the exponent and significand are encoded with a finite number of bits, the minimum step size to the next available value in this schema is a function of the  current value.  Thus, differences between pairs of values of different orders will have different precisions \citep{moler2004numerical}.  For example, when encoding values of order 1 as double precision floats, the minimum step size to the next available value is on the order of 1$\times10^{-16}$.  However, when encoding values of order 1e6 (like, for example, the value of 10 pc in AU), the step size to the next available value is of order 1$\times10^{-10}$.

\begin{figure}[ht]
\centering
\includegraphics[width=0.75\columnwidth]{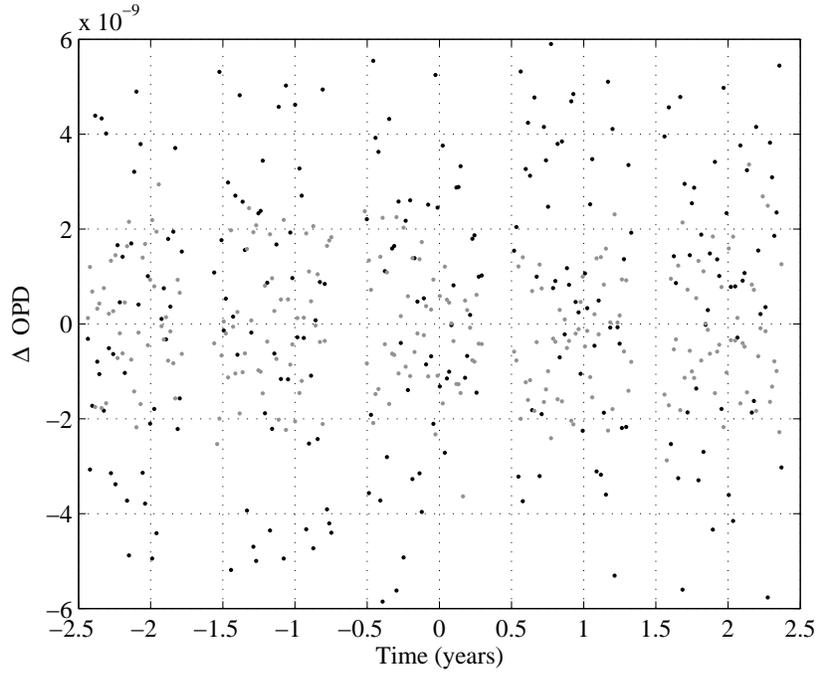}
 \caption[]{Difference between differential OPDs (between a sun-twin star and fixed centroid) using the exact formulation of $\hat\R_{s/sc}$ from equation (\ref{eq:rhat_ssc}) with the default double-precision data type and a multiple precision data type at 256 bits.  All values are the same as in previous simulations, except that the target star is placed at 100 pc. The two curves represent the measurements along two orthogonal interferometer baseline orientations. \label{fig:precision_test}}
\end{figure}

This becomes important when one wants to generate `ground truth' data, using, for example, the exact representation of the astrometric observation in equation (\ref{eq:rhat_ssc}).  In this case, certain values, such as $\R_s(t_0)$ are multiple orders of magnitude greater than 1, whereas the final result is many orders of magnitude less.  This means that numerical noise at the level of the desired signal could inadvertently be introduced into the simulated data.  In the simulations presented here, this only became significant when converting $\R_{s/sc}$ to a unit vector. One solution (the one employed here) is to generate the large values with an arbitrary precision data type using the appropriate number of digits.  Once the spacecraft to star unit vector is generated, all values are of sufficiently low order that the default fixed precision data type is sufficient for the remaining calculations.  To do this, it is necessary to use an arbitrary-precision data type, which simply encodes values using variable length significands, rather than the fixed bit scheme described above.  For the simulations here, this was done using the GNU multiple precision library (\url{http://gmplib.org/}), with the significand stored as 256 bits.  The numerical effect is quite limited in the simulations above (with the target star at 10 pc), but becomes much more pronounced for targets at 100 pc.  Figure \ref{fig:precision_test} shows the difference between differential OPDs using equation (\ref{eq:rhat_ssc}), with one simulation done using the default double precision data type for all values, and the other using the multiple precision data type.  The error between the two is close to 1$\times10^{-8}$ in magnitude, making it a significant noise contribution.

An alternative to this approach would be to use equation (\ref{eq:rhat_norm}).  All terms in this formulation (which is equivalent to equation (\ref{eq:rhat_ssc})) are of order 1 or less, so that the default double precision data type should be sufficient.  However, use of this equation can also introduce numerical noise, if one is not careful in how the various terms are defined.  Terms such as $\R_\mu$ are essentially small values divided by very large values.  If they are not exactly defined, but rather calculated in software, then they will have the same problems as shown above.  Since this discussion applies only to the simulation or generated data, it is relatively simple to just assume values for barycenter motions that will give exact fractional values when scaled by the parallax, eliminating such concerns.

Finally, if we wished to update our expansions from \S\ref{sec:expansion} such that the residual was below the precision of the default double precision data type, we would need to retain terms proportional to $\varpi^3$.  However, in the case of the expansions assuming prior knowledge of parallax and barycenter motion, no higher order terms in $\delta \varpi$ would be needed.

\section{Conclusions}
In this treatment, we have presented an exact vector formulation of a narrow-angle astrometric observation incorporating the effects of parallax, stellar reflex, and barycenter motion.  Because the derivation leading to equations (\ref{eq:rhat_ssc}) and (\ref{eq:rhat_norm}) does not represent a significant computational burden for modern computers, we believe that they (or something fairly similar) should be used for the simulation of data for the purpose of testing analysis techniques, rather than any linearization of any order of precision.  While it may be argued that using a simplified data set (i.e., one derived from approximate or linearized descriptions of the true observation) allows for initial testing of an analysis method, which can then be followed up with more refined tests, we do not believe that such experiments are scaleable.  The most important point to keep in mind when using astrometry for planet finding is that the expected signals will be of very low order, and will interact non-linearly with several unknown (or partially known) parameters.  Therefore, testing data produced by a linearization which may itself introduce residual signals close to the order of the planet signal, and which adds the planet signal without modeling its interaction with other terms, does not actually tell us anything about our analysis technique's ability to isolate the planet signal in a real data set.

Another argument which is often put forward is that such considerations are unimportant because the measurement noise will generally be higher than the errors discussed here.  However, most analysis techniques (be they Bayesian inference or nonlinear programming minimization), make some assumptions as to the structure of the noise.  Even if the noise is not described as additive white Gaussian, any autocorellation and non-zero mean components are carefully modeled based on our knowledge of the physics of the observed systems and instrument.  Linearization residuals are not random and often introduce patterns that can be quite similar to planet signatures, especially due to the parallax effect of any sun-orbiting observatory. Use of the exact observation description obviates the need for all such considerations.  

Furthermore, the linearization presented in \S\ref{sec:expand_apriori} allows us to use a sufficiently precise linear expression for analysis, with only the assumption that we have measurements of parallax and barycenter motion of certain fidelity.  This should be very helpful for any methods reliant on fitting techniques, as the nonlinearities in the second order expansion from \S\ref{sec:expand_noapriori} are quite difficult for most least-squares algorithms.  Again, both the functional representation and computational requirement of this linearization are only slightly more demanding than those of the classical first order astrometric equations, and so there appears to be no reason not to use representations of the astrometric observation derived in a way similar to the one shown here.

\bibliographystyle{plainnat}
\bibliography{Main} 

\begin{table}[ht]
\caption{The vectors used to determine the astrometric measurement.\label{tbl:vectors}}
\vspace{1ex}
\begin{tabular}{l  l}
\tableline\tableline
Symbol & Definition\\
\tableline
$\R_G(t_0)$ & Vector from solar system barycenter to $G$ at epoch $t_0$\\
$\R_{sc}$ & Vector from solar system barycenter to spacecraft at time $t$\\
$\R_{pm}$ & Vector from solar system barycenter to $G$  at time $t$ (movement due to barycenter motion)\\
$\R_{ppm}$ & Vector from spacecraft to $G$ at time $t$ (movement due to barycenter motion and parallax)\\
$\R_{\mu}$ & Vector from $G(t_0)$ to $G$ at time $t$ (barycenter motion) \\
$\R_{s/G}$ & Vector from $G$ to the star at time $t$ (equal to zero if star has no planets)\\
$\R_{s}$ & Vector from solar system barycenter to the star at time $t$ \\
$\R_{c/sc}$ & Vector from spacecraft to centroid of reference stars  \\
$\R_{s/sc}$ & Vector from spacecraft to star at time $t$ (the measurement vector) \\
\tableline
\end{tabular}
\end{table}

\begin{table}[ht]
\caption{Typical values for terms in equation (\ref{eq:rhat_ssc}).\label{tbl:typ_vals}}
\begin{tabular}{ c c l}
\tableline\tableline
Term & Typical Order & Notes \\
\tableline
$a$ & 1 & All distances are assumed to be in AU\\
$\hat\R_s(t_0)$ & 1 &\\
$\bar\R_\mu$\tablenotemark{*} & 5$\times10^{-6}$/year & Barnard's star is 5$\times10^{-5}$/year \\
$\Delta \tilde\R_{s/G}$ & 1$\times10^{-6}$ to 1$\times10^{-2}$ & For solar system planets \\
$ \tilde\R_{sc}$ & 1 & Assuming a ground-based, Earth orbiting or Earth-trailing spacecraft\\
$\varpi $& 5$\times10^{-6}$ to 5$\times10^{-8}$ & For target distance of 1-100 parsecs.\\
\tableline
\end{tabular}
\tablenotetext{*}{While $\bar\R_\mu$ is a time dependent value, for measurements made over a span of 5 to 10 years it will change by less than one order, so we will assume a conservative, constant value when considering which terms to keep in our expansions.}
\end{table}

\end{document}